\documentstyle{lamuphys}
\input psfig.sty
\makeatletter
\let\chapter\hid@chapter
\makeatother
\begin{document}
\pagenumbering{arabic}
\title{Abundance Gradients in the Galactic Disk: a clue to Galaxy Formation}

\author{Francesca\,Matteucci\inst{1}, Cristina\,Chiappini\inst{2}}

\institute{Dipartimento di Astronomia, Universita' di Trieste,
Via G.B. Tiepolo 11,
I-34100 Trieste, Italy and SISSA/ISAS, Trieste, Italy
\and
Departamento de Astronomia,
Observatorio Nacional (CNPq/ON), Rua General Jos\'e Cristino 77,
Sao Cristov\~ao, 20921-030 Rio de Janeiro, Brazil}

\maketitle

\begin{abstract}
We review the observational evidences and the possible theoretical 
explanations for the abundance gradients in the Galactic disk. 
In particular, we discuss the
implications of abundance gradients and gradients of abundance ratios on the
mechanism for the formation of the Galaxy. 
We conclude that an {\bf inside-out} formation of the Galaxy, and 
in particular of the Galactic disk,
where the innermost regions are assumed to have formed much 
faster than the outermost ones, represents the most likely explanation 
for abundance 
gradients and we predict that
the abundance gradients along the Galactic disk have increased with time.   

\end{abstract}

\section{Introduction}

The existence of abundance gradients along the Galactic disk seems now to be
well established with agreement between different data sources.

\begin{itemize}
\item {\bf HII regions}: Shaver et al. (1983) first measured abundance 
gradients from Galactic 
HII regions 
indicating that
O/H, N/H, Ar/H show gradients of the order of $-$0.07 dex Kpc$^{-1}$ between 
4 and 14 Kpc along the Galactic disk. 
More recently,
Vilchez and Esteban (1996) found a flattening of O/H, N/H and S/H gradients for
$R_{GC} > 12$ Kpc suggesting that these gradients present a bimodal behaviour.

\item {\bf Planetary Nebulae}: Maciel and Chiappini (1994) and 
Maciel and Quireza (1998) found 
O/H, Si/H, Ne/H, Ar/H gradients from planetary nebulae very similar 
to those found from
HII regions, namely between $-$0.04 -- $-$0.07 dex Kpc$^{-1}$ in the
galactocentric distance range
4 - 14 Kpc.
 
\item{\bf B type stars}: Kaufer et al. (1994) found that B 
type stars did not exhibit any appreciable gradient but more recent studies (Smartt and Rolleston, 1997 and Gummersbach et al. 1998) found a gradient of oxygen of $-$0.07 $\pm$ 0.01 dex Kpc$^{-1}$ between 6 and 18 Kpc and $-$0.07$ \pm$ 0.02 dex Kpc$^{-1}$ between 5 and 14 Kpc, respectively. These results agree with 
the gradient estimated by Shaver et al. (1983) from HII regions 
and with the planetary nebula data.
\item{\bf Open clusters}: as indicated in the review by Friel (1995)
the iron abundance in open clusters shows
a gradient of [Fe/H]  of $-$0.095$\pm$ 0.017 dex Kpc$^{-1}$ between 7 and 16 Kpc
but Friel (1998) showed a more recent determination of the [Fe/H] gradient shallower than the previous one and closer to the O gradients as estimated recently from B type stars .
\end{itemize}
\par
It is therefore encouraging that different data sources seem to agree at least on the value of the oxygen gradient. However, we would be careful in concluding that
the gradients of all the other heavy elements (C, N, Fe, Ar. etc..) should be the same as oxygen since from a theoretical point of view we would expect small differences due to the different nucleosynthetic production mechanisms
and different progenitors of the elements, as we will see in the next paragraphs.
\par
The question we should ask now is : {\it how can we obtain abundance gradients
along the Galactic disk?}
\par
Several mechanisms have been proposed in the last years:
\begin{itemize} 
\item i) Variations in the gas fraction as a 
function of the galactocentric distance $R_{GC}$ (original idea of the ``simple model'') but 
the predicted gradients are too flat compared to observations (Pagel 1989).

\item ii) Variations in the initial mass function (IMF) but most of 
the suggested variations
(massive stars increase with decreasing metallicity Z) 
tend to destroy the gradient!
Carigi (1996) tried the opposite case, where low mass stars increase with 
decreasing metallicity, and showed that steep abundance gradients 
can be obtained but that the same model does not fit 
the chemical
evolution of the solar neighbourhhod. However, before concluding that 
there is no viable solution
with a variable IMF we should try to test the variation of the IMF 
recently proposed by Larson (1998) who suggested a variation in the lower mass 
cut-off with galactocentric distance rather than a variable slope.

\item iii) Metal dependent stellar yields a la Maeder (1992)
(less metals are produced at higher Z) produce shallow gradients
along the Galactic disk, as it has been shown by several papers 
(Giovagnoli and Tosi,
1995; Carigi 1996), which do not agree with the most recent 
observational estimates.

\item iv) Variations in the
efficiency of star formation ($SFR= \nu(R_{GC}) \sigma_{gas}^{k}$).
Prantzos and Aubert (1995) explored this possibility by assuming
$\nu \propto R_{GC}^{-1}$ and found shallow gradients, again too shallow if compared with the most recent data.

\item v) Variations of the star formation rate (SFR) relative to the infall rate (IR), namely  SFR/IR,  with $R_{GC}$. This hypothesis is also known as  
{\bf biased infall} since one can obtain such a situation by assuming that the Galactic disk formed by infall of gas occurring at a faster rate in the innermost regions than in the outermost ones. The physical reason for 
a biased infall can be found in the fact that the gas tends to collapse more 
quickly in the center of
the spheroid so that a gradient in the gas density, being lower 
in the more external parts, is established. 
In this situation the gas which continues to infall towards the 
disk falls more efficiently towards the center than towards the 
external regions, due to the stronger tidal force acting at the center 
(see also Larson 1976).
Several authors showed that the biased infall can well reproduce steep 
abundance gradients along the disk, especially if coupled with a star 
formation rate proportional to some power $k$ of the gas density
$SFR \propto \sigma_{gas}^{k}$ 
with $k > 1$
(Tosi 1988; Matteucci and Fran\c cois 1989; Matteucci et al. 1989; 
Chiappini et al. 1997). Therefore, we consider this as a viable solution
to explain the existence of abundance gradients.

\item vi) Radial flows along the Galactic disk are not enough by 
themselves to produce steep gradients but they work well if coupled with 
biased infall or a threshold in the SFR under specific conditions
(Clarke 1989; Sommer-Larsen 
and Yoshii 1989; Goetz and Koeppen
1992; Chamcham and Tayler 1994, Edmunds and Greenhow 1995).

\item vii) A biased outflow assuming a stronger and earlier wind in 
the external regions than in the internal ones can also produce steep 
abundance 
gradients, as shown for elliptical galaxies by Martinelli et al. (1998).
At this point one could ask : {\it how can we distinguish between biased infall and biased outflow?}. There is a clear way to do it: thru the abundance ratios.
In fact, under the biased infall hypothesis the 
predicted present time [$\alpha$/Fe] ratios tend to decrease with galactocentric distance. This is due to the fact, as shown by 
Matteucci (1992), that if a region of the Galaxy evolves slower than 
another one then, at the same [Fe/H], the slow region will show a lower 
[$\alpha$/Fe] ratio, due to the longer timescale required 
to achieve 
the same [Fe/H] relative to the fast region and therefore to the larger 
contribution from
type Ia supernovae which are responsible for the bulk of iron production
(Greggio and Renzini, 1983; Matteucci and Greggio 1986), at that given [Fe/H].
On the other hand, under the biased outflow hypothesis 
Martinelli et al. (1998)
predicted that the [$\alpha$/Fe] ratio increases with 
the galactocentric distance, 
due to the fact that the galactic wind occurs before in the external 
regions than the internal ones, and that it is likely that no more 
or negligible 
star formation has taken place after the wind event. As a consequence 
of this, the stars in the external regions will
not show the iron contributed by supernovae of type Ia but rather that 
produced by massive stars. 
In order to decide which one is the mechanism responsible for 
the abundance gradients in the Galactic disk one should check the gradient of
the
[$\alpha$/Fe] ratio.
\end{itemize}

In the following sections we will describe examples of biased infall 
models for the chemical evolution of the Galactic disk and compare their 
predictions with the available data.
In particular, we will discuss abundance gradients of several chemical elements
and of their ratios and by comparing these results with observations we 
will try to infer some conclusions on the mechanism of formation of the Galaxy.

\section{Chemical Evolution Models}

The Matteucci and Fran\c cois (1989)(hereafter MF) model
assumes that the Galactic disk is divided in several independent rings 
2 Kpc wide with no exchange of matter between them.
It assumes also that the disk forms out of infall of primordial gas
which has accumulated with different timescales at different 
galactocentric distances. The galactic halo represents the earliest phase 
of this infall process and no clear distinction between halo and disk 
is present.
In particular, the assumed law for the disk formation is 
a linear relation between $\tau$ and $R_{GC}$:
\begin{equation}
\tau(R_{GC})=
0.464R_{GC}-1.59
\end{equation}

It is worth mentioning that
a quadratic or exponential law for $\tau$ does not produce 
relevant differences, as shown by Carigi (1996), in the predicted abundance gradients. 
The adopted stellar yields are independent of metallicity and the 
star formation rate is proportional to both the surface gas density 
and the total surface mass density, SFR $\propto \sigma_{gas}^{k} 
\sigma_{tot}^{h}$.
Matteucci and Fran\c cois concluded that a good fit to the
observed abundance gradients (Shaver et al. 1983)
and gas distribution in the disk requires
k=1.1 and h=0.1.
\par
Chiappini's et al. (1997) (hereafter CMG) model is a two-infall model, in the
sense that assumes that the halo and the thick disk formed much more quickly 
than the thin disk during a first infall episode,
whereas the thin disk formed by means of a second completely 
independent episode of primordial infalling gas.
In particular, the timescale for the formation of the halo-thick disk is
$\tau_{H} \sim 1$ Gyr and does not vary significantly with the 
galactocentric distance whereas the timescale for the thin disk 
($\tau_{D}$) varies as:

\begin{equation} 
\tau_{D}(R_{GC})=0.875R_{GC}-0.75
\end{equation}

In this approach the evolution of the halo-thick disk is completely
disentangled from that of the thin disk.
The star formation rate is the same as in MF
except that $k=1.5$ and $h=0.5$. 
Is is worth noting that the higher $k$ value adopted by CMG together with 
the longer timescale for the 
thin disk formation (8 Gyrs for the solar vicinity) ensures a good
fit of the reviewed G-dwarf metallicity distribution by Rocha-Pinto 
and Maciel (1996).
Moreover, this choice for the star formation exponent is in agreement with 
recent estimates of 
the star formation rate in spirals by Kennicutt (1998).
This model also adopts
a threshold surface gas density ($\sigma_{gas}=7 M_{\odot} pc^{-2}$) 
for the star formation rate (Kennicutt, 1989) which naturally produces 
a iatus in the star formation rate between the end of the thick disk phase 
and the beginning of the thin disk phase. Such a iatus seems to be real since
it is observed either in the plot
of [Fe/O] versus [O/H] (CMG) or
in the plot of
[Fe/Mg] versus [Mg/H] (Bernkopf and Fuhrmann, 1998).
As we will see, the assumed threshold is also 
responsible  
for the flatter than MF's gradients
in the inner thin-disk regions and
the very flat gradients 
in the outermost thin-disk regions. 
Chamcham and Tayler (1994) 
also found a bimodal gradient due to the adoption of a threshold in the 
star formation rate.

\subsection{Basic Equations}
The basic equation in both models (MF and CMG) 
are:

\begin{eqnarray}
{d \; G_i(r,t) \over dt}  =
  - \; X_i(r,t) \; \Psi(r,t) \; +
\int_{M_L}^{{M_B}_m}\Psi(r, t-\tau_M) \; (Q_M)_i \; \Phi(M) \; dM
 + \nonumber \\ A
 \int_{{M_B}_m}^{{M_B}_M} \Phi(M_B)
\biggl[\int_{\mu_m}^{0.5} f(\mu) \;
\Psi(r, t-{\tau_M}_2) \; ({Q_M}_1)_i(t-{\tau_M}_2) \; d\mu \biggl] \;
dM_B \nonumber \\
 +
(1-A) \int_{{M_B}_m}^{{M_B}_M} \Psi(r,t-{\tau_M}_B) \;
({Q_M}_B)_i(t-{\tau_M}_B) \; \Phi(M_B) \;
dM_B
\; \nonumber \\ +
 \int_{{M_B}_M}^{M_U} \Psi(r,t-\tau_M) \; (Q_M)_i(t-\tau_M) \; \Phi(M) \; dM
+
(X_i)_{inf} \; {d \; G(r,t)_{inf} \over dt}
\end{eqnarray}

In CMG the infall term is defined as:
\begin{equation}
{d \; G_i(r,t)_{\inf} \over dt} =
 {A(r) \over \sigma (r,t_G)} \;
(X_i)_{\inf} \;
e^{-t/\tau _H} + {B(r) \over \sigma (r,t_G)} \; (X_i)_{\inf} \;
e^{-(t-t_{\max})/\tau _D}
\end{equation}

\noindent
where
$G_i(r,t)_{\inf}$ is the normalized surface gas density
of the infalling material in the form of the element $i$,
${(X_i)}_{\inf}$ gives the composition of the infalling gas
which is assumed to be primordial.
$t_{\max}$ is the time of maximum gas accretion onto the disk,
and $\tau_{H}$ and $\tau_{D}$ are the timescales for the mass accretion in
the halo and disk components, respectively.
These are the two really free parameters
of the model and are constrained mainly by comparison with the observed
metallicity distributions for halo and disk stars. The $t_{\max}$ value
is chosen to be 2 Gyrs and roughly corresponds to the end 
of the halo-thick disk 
phase.
The quantities $A(r)$ and $B(r)$
are derived by the condition of reproducing the current total
surface mass density distribution in the halo and the disk,
respectively. The current total mass distribution is taken from Rana (1991).
\par
In the previous model of MF the infall 
term was simply:

\begin{equation}
{d \; G_i(r,t)_{\inf} \over dt} =
 {A(r) \over \sigma (r,t_G)} \;
(X_i)_{\inf} \;
e^{-t/\tau} 
\end{equation}

with a unique $\tau$ for the halo and the disk and where $A(r)$ was 
chosen under the condition of reproducing the present time total surface 
mass density in the disk. In both the models the adopted IMF is taken from Scalo (1986).

\section{Model Predictions}
A good model for the chemical evolution for our Galaxy should honour
a minimal number of observational constraints both in the solar 
neighbourhood and in the whole disk. Both the models described above 
were tested to reproduce
the following constraints in the solar vicinity: 
i) the relative number of disk and halo stars in the solar cylinder, 
ii) the present day gas fraction, iii) the type Ia and II supernova 
rates and their 
ratio, iv) the solar abundances, v) the age-metallicity relation, vi) 
the present day infall rate, vi) the G-dwarf metallicity distribution, 
vii) the [el/Fe] 
versus [Fe/H] relations.
Under these conditions MF concluded that 
good abundance gradients in agreement with the Shaver et al. (1983) data, 
which agree with the most recent estimates, are reproduced by assuming an 
{\bf inside out } formation of the disk, as described by equation (1). 
In addition, 
they showed that the star formation law should have  $0< k \le 1.5$
with a best value of $k=1.1$ in order to reproduce, at the same time,
abundance gradients and an acceptable distribution of the surface 
gas density along the disk. 
In fact, the gas distribution along the disk strongly depends on the
assumed power of the surface gas density in the star formation law.
The predicted gradients, in the galactocentric distance range 4-14 Kpc,
are:
${\Delta log(O/H) \over \Delta R_{GC}}= -$0.065 dex Kpc$^{-1}$;
${\Delta log(^{12}C/H) \over \Delta R_{GC}}= -$0.066 dex Kpc$^{-1}$;
${\Delta log(^{14}N/H) \over \Delta R_{GC}}= -$0.085 dex Kpc$^{-1}$;
${\Delta log(Fe/H) \over \Delta R_{GC}}= -$0.07 dex Kpc$^{-1}$;
${\Delta log(^{4}He/H) \over \Delta R_{GC}}= -$0.0085 dex Kpc$^{-1}$.
As we can see, most of these gradients are in agreement with the most 
recent estimates. An interesting fact is that gradients of different elements 
are slightly different and this is due to the different nucleosynthesis 
history of each element.
For example $^{14}$N, which is mostly produced in low and intermediate 
mass stars as a partly primary and partly secondary (proportional to the 
original abundances of O and C) element, shows a steeper gradient than 
oxygen which is mostly produced in massive short lived stars.
In addition, the fact of having a secondary nature plays a role in defining 
the absolute value of the abundance gradient.
The gradient of Fe is also slightly steeper than that of oxygen since iron 
is mostly produced on long timescales by type Ia SNe (white dwarfs in binary 
systems). Therefore, both the timescales and the primary/secondary nature 
of a chemical element play a role in defining its gradient along the 
Galactic disk.

\par
CMG predicted the existence of bimodal gradients along the Galactic disk, in the sense that their gradients are flatter in the 
outermost regions and steeper in the internal ones ($R < 10$ Kpc), 
in agreement 
with Vilchez and Esteban (1996). In particular, their gradients, 
in the range 4-14 Kpc, are:

\medskip\noindent
${\Delta log(O/H) \over \Delta R_{GC}}= -$0.032 dex Kpc$^{-1}$ in the inner region
and ${\Delta log(O/H) \over \Delta R_{GC}}= -$0.018 dex Kpc$^{-1}$ in the outer region,
${\Delta log(^{14}N/H) \over \Delta R_{GC}}= -$0.037 dex Kpc$^{-1}$ in the inner
region and ${\Delta log(^{14}N/H) \over \Delta R_{GC}}= -$0.028 dex Kpc$^{-1}$
in the outer region, 
${\Delta log(Fe/H) \over \Delta R_{GC}}= -$0.04 dex Kpc$^{-1}$ in the inner
region and 
${\Delta log(Fe/H) \over \Delta R_{GC}}= -$0.017 dex Kpc$^{-1}$ in the 
outer region.

The gradients are all flatter than those predicted by 
MF and the main reason resides in the fact that CMG adopted  a threshold 
surface gas density below which the star formation is assumed to stop.
This mechanism acts mostly in the external regions where the amount 
of gas is almost always close to the threshold.
However, the best model of CMG predicts a gas distribution 
along the disk (see figure 2) in
better agreement than that predicted by MF and
this is also due to the assumed threshold in the star formation rate.

\subsection{Abundance gradients and variable IMF}
An alternative explanation for the formation of abundance gradients 
is a variable IMF, as discussed in the introduction.
In figure 1 we show the predicted gradient of oxygen by the best model of 
CMG (Model A), by a model considering an IMF which predicts more 
massive stars in lower density regions (Padoan et al. 1997) (Model B)
and the predictions of Model C where both the variable IMF and the 
variable star formation efficiency are considered.
The star formation efficiency is assumed to increase with decreasing
$R_{GC}$ as in Prantzos and Aubert (1995).
It is clear that a variable IMF of this kind tends to destroy the gradient unless the additional hypothesis of the variable star formation efficiency is considered.
However, as shown in figure 2, this last model (Model C) does not 
fit at all the observed gas distribution along the disk and 
therefore it should be rejected.


\begin{figure}
\centerline{\psfig{figure=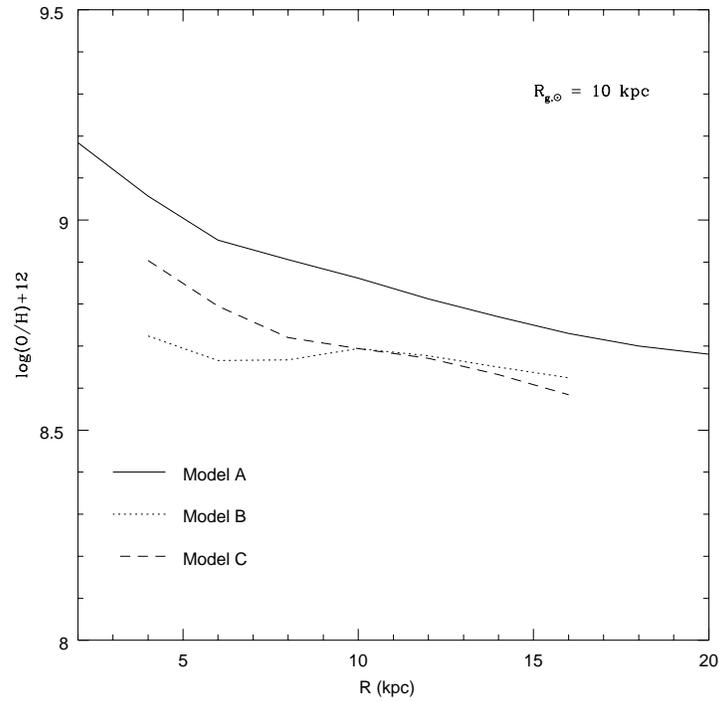,height=10cm,angle=0} }
\caption{The predicted oxygen abundance along the Galactic disk 
by three different models}
\end{figure}

\begin{figure}
\centerline{\psfig{figure=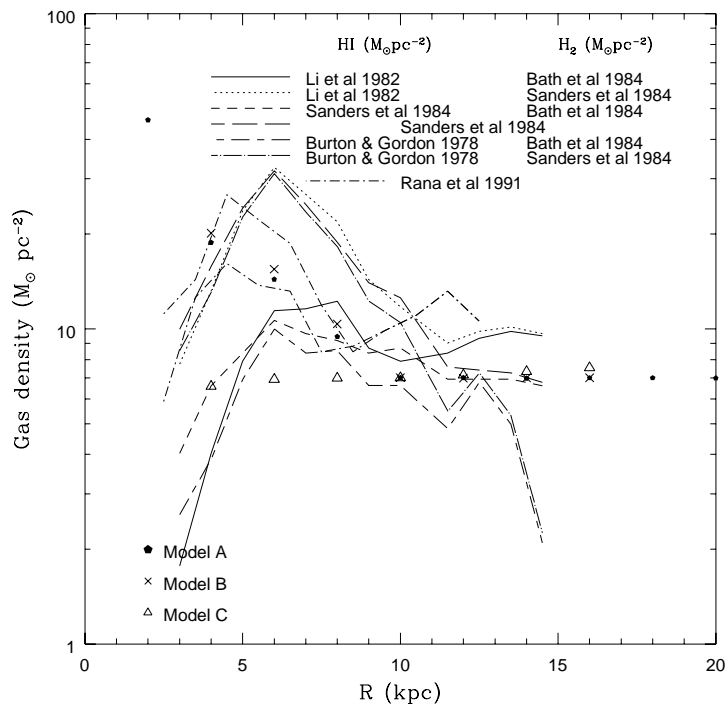,height=10cm,angle=0} }
\caption{Observed and predicted gas distribution along the Galactic disk.
The data and the models are indicated  in the figure.}
\end{figure}

\section{ Discussion and conclusions} 

In this paper we have shown that the most realistic scenario to explain 
abundance gradients along the Galactic disk is the {\bf inside-out} 
formation of the disk, in the
sense that the internal parts of the disk are assumed to have formed 
on shorter timescales than the external parts. This allows us to 
reproduce gradients in agreement with the observations as long as 
the exponent of the star formation law,
$k$, is larger than unity. At the same time we can reproduce the 
observed gas distribution along the disk as long as $k \le 1.5$. 
Therefore, we can envisage 
a range of possible values for $k$, namely $1.1 \le k \le 1.5$.
Better agreement is obtained if also a dependence on the total surface 
mass density is considered in the star formation law (see also Tosi 1988).
\par
Concerning the gas distribution along the Galactic disk, we showed 
that the assumption of a threshold in the star formation rate improves the 
agreement with observations although it predicts slightly too shallow 
gradients, 
at least in the light of the more recent data.
\par
The assumption of a variable IMF where massive stars are more 
abundant in low metallicity, low density regions tend to destroy the gradient 
along the disk, at variance with observations, although we can not exclude
the existence of a particular variation of the IMF which can reproduce all the main observational constraints.
\par 
{\it Therefore, our main conclusion is that an inside-out scenario with 
constant IMF
is the most favoured for the formation of abundance gradients}.
\par
In order to check this point we should try to observe [$\alpha$/Fe]
ratios in stars in the outermost disk regions. In fact,
we have also shown that one should expect differences (although small) 
between the gradients of elements produced on different timescales.
For this reason we expect that the [$\alpha$/Fe] ratio has 
a {\bf negative} gradient with increasing
$R_{GC}$, and future measurements of such a ratio at large galactocentric 
distances can assess this point.
Recently, Nissen and Schuster (1997) 
identified metal rich halo stars ([Fe/H] $\sim$ -1 dex)
with both high and low [$\alpha$/Fe] ratios. The halo stars with 
low [$\alpha$/Fe] ratios tend to be on orbits biased to the outer halo. 
This could be the consequence of the inside-out Galaxy formation, in the sense that the outer halo formed more slowly than the inner one in analogy with what happens in the disk. In particular, we predict that the stars 
in the outer disk should show this effect even more than the outer halo stars.
\par

Another important fact, strictly related to the mechanism of Galaxy 
formation, is the evolution of gradients with time. 
On this point there is no general agreement
between different data sources and different authors!

From the observational point of view the situation can be summarized 
as follows:
\begin{itemize}
\item {\bf Planetary nebulae} seem to suggest that  gradients steepen 
with time (Maciel and Koeppen 1994); in fact, 
type III PNe
(the oldest) have a flatter gradient than type II and I PNe, but
dynamical effects could perhaps be responsible for this behaviour.

\item {\bf Open Clusters} suggest no clear variation of gradients 
with time, and 
may be a flattening (Carraro et al. 1998), but again dynamical 
effects could have played a role.
\end{itemize}

From the theoretical point of view some models
predict a steepening of the gradients with time (Tosi 1988;
MF; Goetz and Koeppen 1992; CMG)
whereas others predict the opposite trend
(Ferrini et al. 1994; Allen et al. 1998).
Generally, models with biased infall (inside-out formation of the disk),
such as those discussed here, predict a steepening of gradients in time
due mainly to the assumed long infall timescales at large 
galactocentric distances (see also the review of Tosi 1996).
Clearly the evolution of gradients with time is strictly related to 
the mechanism of Galaxy formation and is an important point to assess. However,
given the situation, it is still premature to draw firm conclusions 
and more precise observational data are necessary.

%
%
%
%

\end{document}